\documentclass[pt=11,twocolumn,preprintnumbers,superscriptaddress, showkeys]{revtex4-2}
\usepackage[dvipsnames]{xcolor}
\usepackage{hyperref}
\usepackage{physics}
\usepackage{graphicx}
\usepackage{amsmath}
\usepackage{float}
\usepackage{amssymb}
\usepackage{bm}
\usepackage{mathtools}
\usepackage{xspace,slashed}
\usepackage{ dsfont }
\usepackage[OT1]{fontenc}
\usepackage{cmbright}
\DeclareFontShape{OT1}{cmss}{m}{it}{<->ssub*cmss/m/sl}{}

\usepackage{xcolor}
\usepackage{braket}

\begin{document}

\title{Determining probability density functions with adiabatic quantum computing}

\preprint{TIF-UNIMI-2023-9, CERN-TH-2023-042}

\newcommand{\MIaff}{TIF Lab, Dipartimento di Fisica, Universit\`a degli Studi
  di Milano, Milan, Italy}

\newcommand{\INFNaff}{INFN, Sezione di Milano, I-20133 Milan, Italy}

\newcommand{\TII}{Quantum Research Center, Technology Innovation Institute, Abu Dhabi, UAE}

\newcommand{\CERNaff}{CERN, Theoretical Physics Department, CH-1211
  Geneva 23, Switzerland}

\newcommand{\CERNgen}{European Organization for Nuclear Research (CERN), Geneva 1211, Switzerland}

\newcommand{\Qibo}{\texttt{Qibo}\xspace}

\author{Matteo Robbiati}
\affiliation{\CERNgen}
\affiliation{\MIaff}
\author{Juan M. Cruz-Martinez}
\affiliation{\CERNaff}
\author{Stefano Carrazza}
\affiliation{\MIaff}
\affiliation{\INFNaff}
\affiliation{\CERNaff}
\affiliation{\TII}

\begin{abstract}

The two main approaches to quantum computing are gate-based computation and 
analog computation, which are polynomially equivalent in terms of complexity, 
and they are often seen as alternatives to each other. 
In this work, we present a method for fitting one-dimensional probability distributions 
as a practical example of how analog and gate-based computation can be used 
together to perform different tasks within a single algorithm.
In particular, we propose a strategy for encoding data within an adiabatic evolution 
model, which accomodates the fitting of strictly monotonic functions, as it is the 
cumulative distribution function of a dataset. Subsequently, 
we use a Trotter-bounded procedure to translate the adiabatic evolution into a 
quantum circuit in which the evolution time $t$ is identified with the parameters 
of the circuit. This facilitates computing the probability density as derivative of the cumulative function
using parameter shift rules.
\end{abstract}


\maketitle

\section{Introduction}

In the context of quantum computing, we are witnessing the development of various 
technologies, which can be categorized into two different but computationally 
equivalent approaches: Gate-Based Computation (GBC) and analog computation~\cite{Albash_2018} (AQC).

These two approaches are often used to address very different types of problems. 
On one hand, many of the most well-known quantum computing query algorithms such as
Shor's~\cite{Shor_1997}, Grover's~\cite{grover1996fast} or Deutsch-Josza's~\cite{Deutsch1992RapidSO} 
are formalized through the gate computation paradigm. Also in the field of Quantum 
Machine Learning (QML)~\cite{Schuld_2014,Biamonte_2017}, 
the most common approach involves defining Parametric Quantum 
Circuits~\cite{Benedetti_2019, chen2020variational, Cerezo_2021} 
that serve as variational models which are trained to perform the target tasks. On the other hand, 
AQC has been shown to be an effective tool for tackling optimization 
problems~\cite{Yarkoni_2022, zaech2022adiabatic, Pelofske_2023, Date_2021, tasseff2022emergingpotentialquantumannealing}, 
in particular Quadratic Unconstrained Binary Optimization (QUBO) problems,
which can be easily encoded within a system of interacting nearest-neighbours particles 
and represented in terms of Ising Hamiltonians.  

In this work, we present an application where AQC and GBC can be used together
in the context of QML,
addressing different tasks within the process by exploiting their respective strengths. 
To showcase our proposed algorithm we tackle the determination of 
the underlying Probability Density Function (PDF) of a
given one-dimensional dataset.

\begin{figure*}[ht]
  \includegraphics[width=0.85\linewidth]{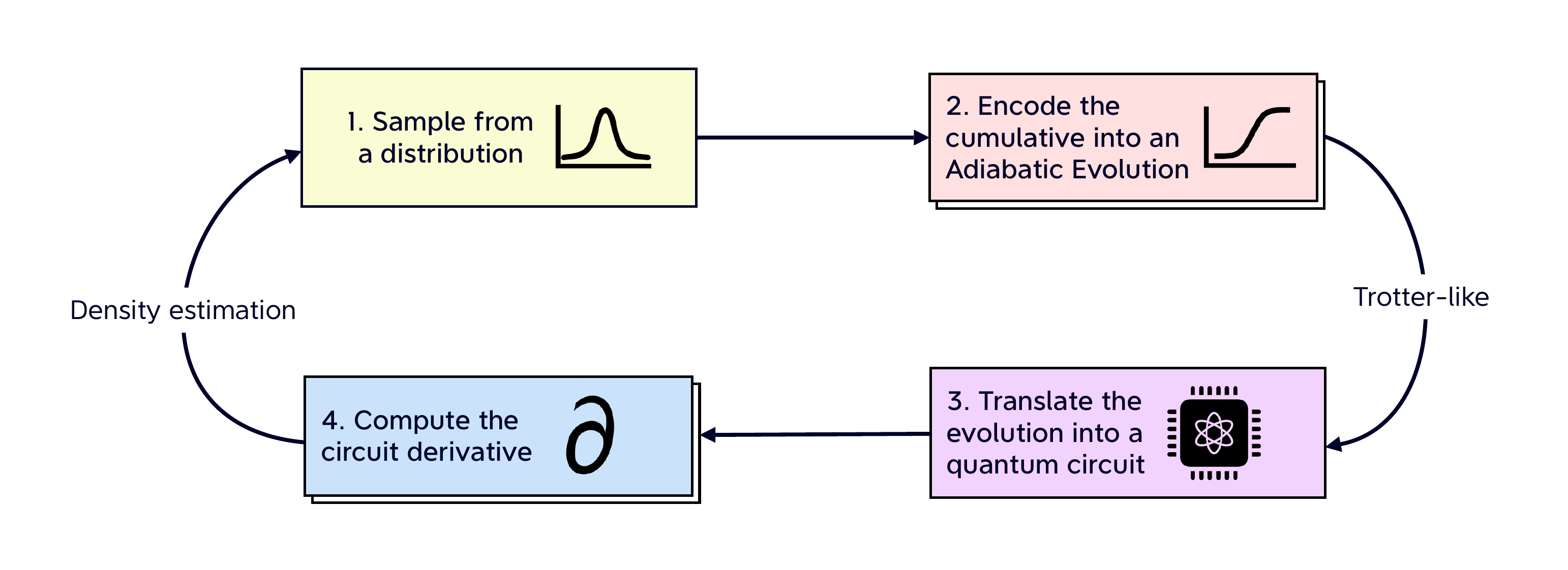}
  \caption{Schematic representation of the proposed Quantum Adiabatic Machine Learning 
  (QAML) algorithm. A cumulative distribution function is encoded
  into an adiabatic evolution, which is then translated into a quantum circuit and 
  differentiated with respect to the target values in order to approximate the probability
  density function.}
  \label{fig:qaml_scheme}
\end{figure*}

This is a problem that presents some very specific challenges. First, given a random 
sample of a distribution, a way of reconstructing the underlying distribution is 
to compute its Cumulative Distribution Function (CDF). An accurate representation 
of a CDF requires a model which behaves monotonically with a target parameter, for which 
we exploit AQC. Then, given its CDF, the PDF can be determined by the derivative 
of the CDF. We face this second challenge through GBC (Fig.~\ref{fig:qaml_scheme}).

Let us then begin by considering the cumulative distribution of a sample. 
We first define a regression model based on adiabatic
evolution~\cite{adiabatic_evolution} which encodes a generic one-dimensional
function defined in a predefined bounded range as the time evolution of the expectation 
value of an arbitrary observable over the evolved state. This
approach is sufficiently flexible to fit a large variety of functional forms and 
it can be easily set up so that boundary conditions and the monotonicity of the 
problem are automatically satisfied with a suitable definition of the adiabatic evolution. 
This final remark is particularly important when 
dealing with Cumulative Distribution Functions (CDF), which have to be monotonically
increasing and defined between $0$ and $1$.

After achieving an acceptable CDF fit, the method projects the obtained adiabatic
Hamiltonian into a quantum circuit representation using a Trotter-like
decomposition~\cite{Paeckel_2019}. This step opens the possibility to train 
and perform inference of the
regression model on circuit-based quantum devices and therefore give us the
possibility to extract the PDF of the sample as the derivative of the
circuit using robust and well known Parameter Shift Rules (PSR)~\cite{Mitarai_2018,Schuld_2019}.

The paper is organized as follows. In Sec.~\ref{sec:methodology} we present the
technical details of the probability density function estimation using using the 
synergistic action of analog and gate-based quantum computing. The Sec.~\ref{sec:validation} presents validation results for
multiple examples and a comparison with classical Kernel Density Estimation (KDE) methods to verify the 
robustness of our algorithm in terms of accuracy. 
Finally, in Sec.~\ref{sec:conclusion} we draw our conclusion
and outlook.

\section{Methodology}
\label{sec:methodology}

In this section we describe the procedure implemented for the determination of
probability density functions. The algorithm is separated in two steps: the
approximation of an empirical cumulative distribution function using adiabatic
quantum evolution in a discrete time-grid as a regression model, and subsequently, the determination of
the probability density function through the trotter-like quantum circuit
representation obtained from the adiabatic Hamiltonian. In the same section we 
generalize the evolution time to a continuous variable.

In Sec.~\ref{sec:regressor} and Sec.~\ref{sec:training} we introduce the choice 
of using an adiabatic evolution to fit a cumulative distribution. And in Sec.~\ref{sec:tocirc} we translate the obtained operator into a quantum circuit, 
facilitating the computation of its derivatives. 

\subsection{Model regression with adiabatic quantum evolution}
\label{sec:regressor}

Given a function $F(t)$, one-dimensional in input and output, we build a regression model by
selecting an observable $\mathcal{O}$ such that there are two Hamiltonians, $H_{0}$ and
$H_{1}$, for which the expectation value of $\mathcal{O}$ over the ground states of 
$H_0$ and $H_1$ correspond to the two
points between which we want to learn the function $F$.

Therefore, we interpret the regression problem as the procedure of building a
time dependent Hamiltonian $H(t)$, such that its ground state $\ket{\psi(t)}$ at time $t$ satisfies
\begin{equation}
  \langle \psi(t) | \mathcal{O} | \psi(t) \rangle = F(t) \\.
  \label{eq:expO}
\end{equation}
From now on, for simplicity, we shorten the l.h.s. of the expression~\eqref{eq:expO} with 
$\braket{\mathcal{O}}_t$.
We construct this Hamiltonian implementing an adiabatic evolution
\begin{equation}
    H(t) = \bigr[1-s(t;\bm{\theta})\bigl] H_0 + s(t;\bm{\theta}) H_1 \\,
    \label{eq:Had}
\end{equation}
governed by the parametric scheduling function $s(t;\bm{\theta})$, where $t$ has to be 
defined in $[0,1]$. 
The problem is then reduced to finding the right set of
parameters $\bm{\theta}$ such that $\braket{\mathcal{O}}_t$ during the 
adiabatic evolution of the state
$\ket{\psi(t)}$ approximates the target function.

Note that the choice of $s(t, \bm{\theta})$ is fundamental to guarantee 
the monotonicity of the target function.

\subsection{Learning empirical cumulative density functions}
\label{sec:training}

\subsubsection{Adiabatic evolution setup}
\label{sec:ae_setup}
The presented framework can be applied to the problem of fitting a cumulative 
distribution function $F(t)$, with $t\in[0,T]$. 
This can be done if two requirements are satisfied: the model has to be 
strictly monotonically increasing in $t$ and the extreme values of the function are set to be $F(0)=0$ and $F(T)=1$.   

The second condition can be fulfilled by appropriately selecting the Hamiltonians 
$H_0$ and $H_1$, as well as the observable $\mathcal{O}$. In particular, since we 
focus on one-dimensional distributions, and we treat here the introductory case of one 
qubit, a proper choice can be $\mathcal{O}=\sigma_z$, $H_{0} = \sigma_x$ and $H_{1} = -\sigma_z$,
where $\sigma_x$ and $\sigma_z$ correspond to the Pauli X and Pauli Z matrices respectively. 
This choice satisfies the boundary conditions of the problem $\braket{\mathcal{O}}_0=0$
and $\braket{\mathcal{O}}_T=1$.
Secondly, the monotonicity of the function can be ensured by implementing a scheduling 
function which is monotonic itself. This final remark, together with the appropriate 
Hamiltonians definition, make adiabatic evolution an extremely effective 
model for approximating a CDF.
In this work, we use as scheduling function a polynomial
of degree $p$:
\begin{equation}
s(t;\theta) = \frac{1}{\eta} \sum_{i=1}^{p} \theta_i t^{i}, \qquad
\text{with} \qquad \eta = \sum_{i=1}^{p} \theta_i \\,
\label{eq:scheduling_ansatz}
\end{equation}
whose monotonicity is guaranteed by requiring positive coefficients $\theta_i$ for $t \in [0,1]$.
If one prefers to define a more flexible scheduling function, such as a neural network, 
another viable option is to penalize the model during training when negative 
fluctuations of the predictions are recorded. 

\subsubsection{Training of the adiabatic evolution}
\label{sec:training}

In order to train a parametric adiabatic evolution to approximate a function 
we make use of \Qibo~\cite{stavros_efthymiou_2023_7606063, stavros_efthymiou_2023_7736837,
stavros_efthymiou_2023_7748527, andrea_pasquale_2023_7662185}, which is an open-source
full-stack framework for quantum computing. 
\Qibo provides the possibility to execute analog quantum computing models that 
can be symbolically defined through an interface based on \texttt{SymPy}~\cite{10.7717/peerj-cs.103}. 
These models can then be translated into circuits and executed through trotterization.
In particular, a second-order Time-Evolving Block Decimation (TEBD) method is implemented (see Eq. (62) of Ref.~\cite{Paeckel_2019}).
This approximation, once chosen a time step $\dd \tau$ presents an error of the order 
$O(\dd \tau^3)$ per time step~\cite{Paeckel_2019}. Considering a total evolution time 
$T$ divided into $N=T/\dd \tau$ steps, the total error
of the approximation becomes $O(\dd \tau^2)$. 
Further details can be found in the official \Qibo documentation~\cite{qibo_trotter}.

The use of \Qibo allows us to execute our algorithm on self-hosted superconducting quantum devices (see Sec.~\ref{sec:hardware}),
however by this choice we are limited to a gate-based paradigm,
which forces us to consider a discretized case of the time evolution.
We address this limitation in Sec.~\ref{sec:circuittimes}, by implementing an approximation 
that allows us to extend from the discrete case to the continuous.
This approximation would not be necessary if using analog devices or simulators.
Among the simulation tools we mention QuTip~\cite{Johansson_2012} and HOQST~\cite{Chen_2022}
for handling open quantum systems, and \texttt{QuantumAnnealing.jl}~\cite{morrell2024quantumannealing} for quantum annealing algorithms.  

The procedure follows a supervised machine
learning strategy: at first, we generate a sample of random variables $\{x\}$
following a chosen distribution $\rho$, and we calculate the empirical CDF of the sample
$\{F\}$. This step is equivalent to constructing a cumulative histogram of the data, 
where we select $N_{\rm train}$ bins identified with the evolution times controlling 
the scheduling function. Each 
pair of $\bigl(x_j, F_j\bigr)$ values are mapped into $\bigl(\tau_j, \braket{\mathcal{O}}_{\tau_j}\bigr)$, where $\tau_j=t_j/T$ 
is the evolution time corresponding to the $j$-th step of the discretized adiabatic 
evolution normalized with respect to the final time $T$ and $\braket{\mathcal{O}}_{\tau_j}$ 
is the expectation value of the target observable over the ground state of $H(\tau_j)$
as introduced in Eq.~\eqref{eq:expO}.

In this last step, we apply a transformation to the domain of the variable $x$, 
so that both $x$ and $\tau$ are now defined in the interval $[0,1]$. 
This allows us to define a standardized framework for each analyzed variable. 
The original distribution can be easily reconstructed by applying the inverse transformation.

Once the training set is defined and the scheduling function is initialised using an
initial set of parameters $\bm{\theta}_0$, we execute the adiabatic evolution of the 
state $\ket{\psi(\tau)}$ from $\ket{\psi(0)}$ to $\ket{\psi(1)}$ and we collect 
all the $\braket{\mathcal{O}}_{\tau_j}$ values during the evolution. 

We define a mean-squared error loss function $J$ for estimating the quality of
the fit:
\begin{equation}
  J = \frac{1}{N_{\rm train}} \sum_{j=1}^{N_{\rm train}} \bigl[ F_j - \braket{\mathcal{O}}_{\tau_j}\bigr]^2,
  \label{eq:loss_function}
\end{equation}
where $\braket{\mathcal{O}}_{\tau_j}$ depends on the variational parameters because 
the evolved state on which it is calculated follows an evolution governed by the 
parametric scheduling $s(\tau,\bm{\theta})$. 

We finally perform the training of the model by optimizing the parameters $\bm{\theta}$
using a selected optimizer. In this work we make use of the 
Covariance Matrix Adaptation Evolution Strategy (CMA-ES)~\cite{CMA}, which is 
one of the optimizers provided by \Qibo. We remark that any optimizer can be used 
since the approach is totally agnostic under this aspect.

\begin{figure}[ht]
  \includegraphics[width=1\linewidth]{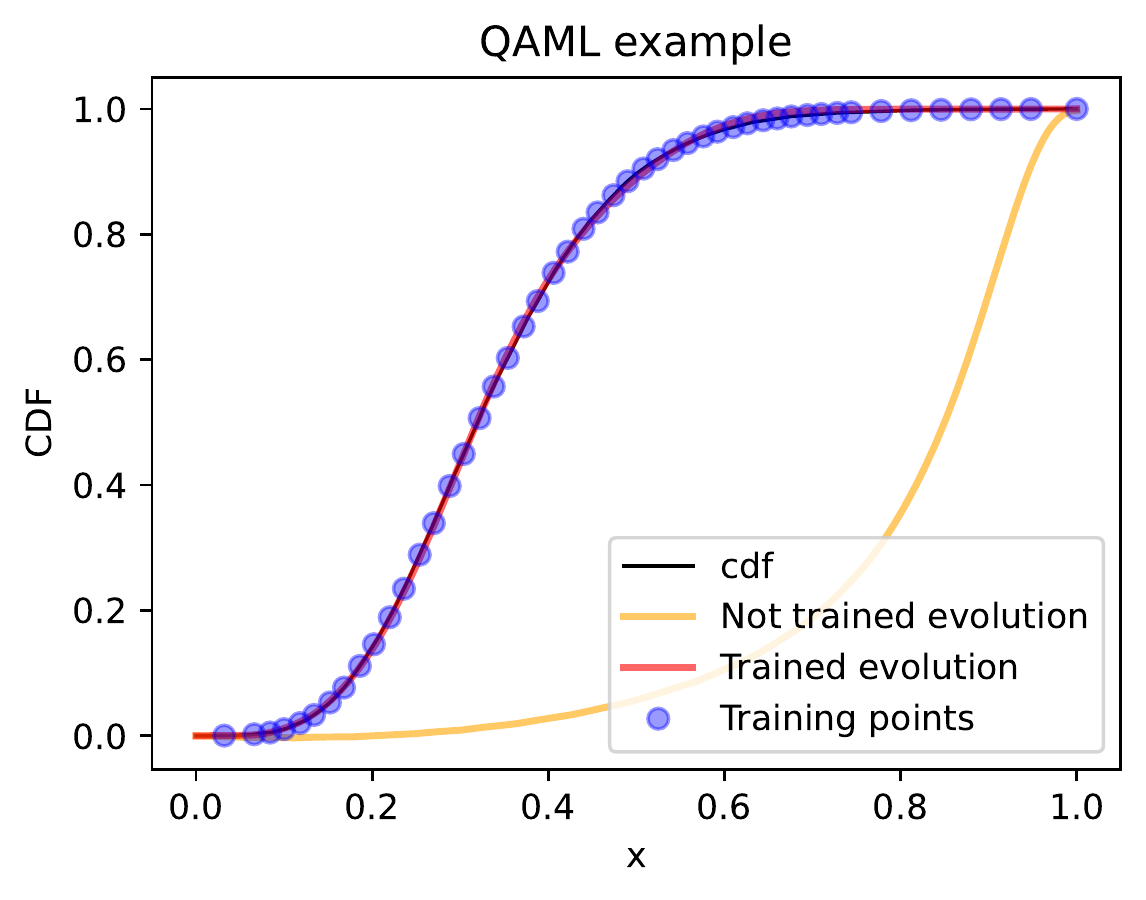}
  \caption{Example of Quantum Adiabatic Machine Learning (QAML). We select 
  $N_{\rm train}$ data from a sample of points picked up from a Gamma distribution. 
  The training labels (blue points) lay on the empirical CDF of the sample (black line).
  The untrained sequence of $\braket{\mathcal{O}}$ (yellow line) is compared with 
  the values after the QAML training (red line). The scheduling function 
  is the one presented in Eq.~\eqref{eq:scheduling_ansatz} with $p=12$.}
  \label{fig:evolution_example}
\end{figure}

In Fig.~\ref{fig:evolution_example} we show and describe an example of the algorithm
presented above, to which we refer from now on as Quantum Adiabatic Machine Learning (QAML).

\subsection{Deriving probability functions from quantum circuits}

In the following we generalize the presented model to accept any continuously sampled 
time $t\in[0,T]$ and we use a quantum circuit representation to compute the 
derivative of the approximated CDF with respect to the target variable. 

\subsubsection{Generalizing the evolution to any time}
\label{sec:circuittimes}

The procedure presented in the previous paragraphs can be interpreted as the
evolution product of a series $\{H(\tau_j)\}$ of Hamiltonians corresponding to the adiabatic
Hamiltonian~\eqref{eq:Had} at fixed evolution time steps $\{\tau_j = t_j/T\}$, discretized
according to $\dd\tau = (t_j - t_{j-1})/T$, the time step of the adiabatic evolution. Each of these
Hamiltonians can be associated to a \textit{local} time evolution operator
$U(\tau_{j})$ which evolves $\ket{\psi(\tau_{j-1})}$ to $\ket{\psi(\tau_j)}$.
More generally, we can obtain any state $\ket{\psi(\tau_n)}$ by sequentially
applying $n$ operators:
\begin{equation}
    \ket{\psi(\tau_n)} = \prod_{j=1}^{n} \mathcal{T} U(\tau_j)\ket{\psi(\tau_{0})} 
    \coloneqq \mathcal{C}(\tau_n)\ket{\psi(\tau_{0})} \\,
    \label{eq:taun}
\end{equation}
where we sum up the sequential action of the $U(\tau_j)$'s into a single
unitary operator $\mathcal{C}(\tau_n)$, which evolves the initial state 
$\ket{\psi(\tau_{0})}$ to the one at time $\tau_n = n \, \dd \tau$ and depends on 
the time-ordering operator $\mathcal{T}$.
From now on, we refer to $H(\tau_j)$, $U(\tau_j)$ and $C(\tau_j)$ as $H_j$, $U_j$
and $C_j$ for simplicity.

In order to compute the state at any value of $\tau$
outside of the discrete simulated time steps of the adiabatic evolution $\{\tau_j\}$
it is necessary to take the continuous limit.
We start by considering one of the intermediate elements of the product introduced 
in Eq.~\eqref{eq:taun}
\begin{equation}
  U_j = e^{-i \text{d}\tau H_j}, \qquad \text{with} \qquad H_j =
  \begin{pmatrix}
    s_j & 1-s_j \\ 1-s_j & -s_j
  \end{pmatrix},
\end{equation}
where $s_j$ is the value of the scheduling at evolution time
$\tau_j = j\,\text{d}\tau$.
This instantaneous form of the adiabatic Hamiltonian operator
can be diagonalized as $D_j$ using a matrix $P_j$ such that
$H_j = P_j D_{j} P_j^{-1}$, where:
\begin{equation}
    P_{j} = \Lambda_{j}
  \begin{pmatrix}
      1 & \frac{s_{j}-\lambda_j}{1-s_{j}} \\ \frac{\lambda_j-s_j}{1-s_j} & 1
  \end{pmatrix},
  \qquad D_j =
  \begin{pmatrix}
    \lambda_j & 0 \\ 0 & -\lambda_j
  \end{pmatrix},
  \label{diagonalization}
\end{equation}
and $\Lambda_j$ is the appropriate ($\tau$ dependent) normalization constant.
The absolute value of the eigenvalues of the Hamiltonian is
$\lambda_j = \sqrt{2s_j^2 - 2s_j + 1}$.

We now use this decomposition to write $\mathcal{C}_n$ in terms of the diagonal form of the Hamiltonian:
\begin{equation}
    \mathcal{C}_n = \prod_{j=0}^{n} P_j e^{-i D_{j}\, \dd\tau } P_j^{-1}. \label{eq:C_with_diagonalization}
\end{equation}

If we now take the limit $\dd\tau \to 0$,
we have $H_j \to H_{j-1}$ and thus $P_{j} \to P_{j-1}$.
We have verified that for smoothly behaved scheduling function adjacent elements in the sequence tend to the identity $P_{j}^{-1}P_{j-1}\to I$ with an error proportional to the time step $\dd \tau$.
The Eq.~(\ref{eq:C_with_diagonalization}) simplifies to
\begin{equation}
    \mathcal{C}_n = P_n \exp{-i \sum_{j=0}^{n} D_j\,\dd\tau} P_0^{-1}.
\end{equation}

Furthermore, in the limit of $\dd\tau \to 0$, $n\to\infty$ and the sum in the above equation becomes an integration
in $\dd\tau$ between the initial and final steps in the evolution.
\begin{equation}
    \mathcal{C}_{t} = P_{t} \exp{-i \displaystyle\int_{0}^{t/T} D_j\,\dd\tau} P_0^{-1},
    \label{eq:Ctau}
\end{equation}
where we now indicate with $P_{t}$ and $P_0$ the diagonalization matrices corresponding
respectively to the last and the first evolution operators we must apply to $H_0$'s
ground state in order to obtain the evolved state at an arbitrary time $t$, dropping the discretization requirement.

The proposed approximation only holds when considering the adiabatic regime, implemented
through a small time step and a slowly varying scheduling function. 
More in general, one can construct a more 
complete and arbitrary accurate representation of $U_t$ exploiting the Magnus Expansion (ME)~\cite{Blanes_2009, Blanes2010APA}. 
Combining a proper choice of the ME order with an arbitrary small trotterization 
error (defined by the Trotter-Suzuki order and step size) one can construct an arbitrarily accurate approximation of $U_t$
as discussed in~\cite{Gonzalez-Conde:2024yfl}. 
% Our derivation corresponds to adopting a first-order Magnus expansion with a single
% second-order TEBD step. 

\subsubsection{Circuit representation}
\label{sec:tocirc}

Let us now implement the unitary operator $\mathcal{C}_{t}$ as a quantum circuit,
using a gate decomposition which is useful for calculating the derivatives of the
circuit with respect to its variational parameters.

To that end we write $\mathcal{C}_{t}$ in terms of rotational gates, with 
the aim of using well-known parameter shift 
rules~\cite{Mitarai_2018, Schuld_2019, Mari_2021} 
on them. The choice of this differentiation 
method relies on its robustness to noise and acknowledging that in the one-dimensional 
case, the decomposition we propose is simple and computationally lightweight. 

Since any unitary operator $\mathcal{U} \in SU(2)$ can be written as a combination of three rotations~\cite{Bertini_2006}
we choose:
\begin{equation}
    \mathcal{U} \equiv R_z(\phi)R_x(\theta)R_z(\psi),  \label{eq:rotation_circuit}
\end{equation}
where the three angles $(\phi, \theta, \psi)$ can be computed as function of the matrix
element of the operator $\mathcal{C}$:
\begin{equation}
    \begin{cases}
        \phi = \pi/2 - \text{arg}(c_{01}) - \text{arg}(c_{00}), \\
        \theta = -2\arccos(|c_{00}|), \\
        \psi = \text{arg}(c_{01}) - \pi/2 - \text{arg}(c_{00}).
    \end{cases}
    \label{eq:euler_angles}
\end{equation}

In~\eqref{eq:euler_angles} the matrix elements $c_{00}$ and $c_{01}$ 
depend on the values of the scheduling $s$ and on the eigenvalues $\lambda$ of $H_{t}$. 
This dependence can be written more explicitly:
\begin{align}
    c_{0j} = \frac{1-s}{s\sqrt{\lambda (\lambda - s)}} & \left\{ \cos{\mathcal{I}}\left(1 + (-1)^{j}\frac{\lambda-s}{1-s}\right) \right. \nonumber \\
           & + \left.  i\sin{\mathcal{I}}\left(1 - (-1)^{j} \frac{\lambda-s}{1-s}\right) \label{eq:c00_and_c01} \right\},
\end{align}
with $\mathcal{I} = \displaystyle\int_{0}^{\tau} \lambda(t')\dd t'$, $s = s(\tau)$, $\lambda = \lambda(\tau)$ and $\tau = t/T$.

Note that the construction of the circuit is completely independent of
the choice of scheduling function. 
Depending on this choice, $\mathcal{I}$ may be computed analytically.

With this, we define a method to build a quantum circuit which, for fixed evolution
parameters $\bm{\theta}$ and time $t$, returns the ground state of $H_{t}$, 
which can be used to compute the requested $\braket{\mathcal{O}}_{t}$.

\subsubsection{From the CDF to the PDF}
\label{sec:cdf2pdf}

The circuit representation of the operator $\mathcal{C}_{t}$ allows
us to reconstruct our original target function
$F(t)$ introduced in Sec.~\ref{sec:regressor} by applying the circuit to a state prepared as
$\ket{\psi(0)}$ and then measuring the chosen observable ($\sigma_z$ in this work) 
over the obtained final state.

As previously said, our example case has been that in which the target function
$F(t)$ correspond to the empirical CDF of 
some arbitrary distribution $\rho$.
By imposing monotonicity and pinning the initial and final points we ensure that its first derivative
corresponds to the PDF of the same distribution.

For a one-dimensional distribution we have then:
\begin{equation}
    \frac{\dd F(t)}{\dd t} =\frac{\dd }{\dd t} \braket{\psi(0) | \mathcal{C}_{t}^{\dagger}\, \sigma_z\, \mathcal{C}_{t}| \psi(0)}.
  \label{eq:derivative_wrt_t}
\end{equation}

In the context of quantum computing, as previously anticipated, 
we can take advantage of what is usually known as Parameter Shift Rule
(PSR)~\cite{Mitarai_2018, Schuld_2019} which allows us to take the derivative of the 
expectation of an observable such as Eq.~\eqref{eq:derivative_wrt_t} by simply evaluating the circuit after shifting the
parameters with respect to which we are taking the derivative.
We are using specifically the formula presented in~\cite{Mitarai_2018}, for circuits based on rotations.
Note that in this case we have limited ourselves to gates in which the parameter appears
only once, but more complicated forms can also be utilized~\cite{Wierichs_2022}.

With this we arrive to the final formula of the PDF in terms of the original circuit for an arbitrary value of $t$:
\begin{equation}
    \rho(t) = \text{PSR}\left[ \braket{\psi(0) | \mathcal{C}_{t}^{\dagger} \, \sigma_z \, \mathcal{C}_{t}| \psi(0)} \right].
\end{equation}

\section{Validation}
\label{sec:validation}

In the following we test the presented algorithm by drawing samples from a known distribution,
building the circuit and reconstructing the original probability function.
All results for this section are summarized in Table~\ref{tab:summary}.

\subsection{Sampling known distributions}

  \begin{table*}[ht]
  \begin{tabular}{lcccccccc}
  \hline \hline
    Fit function & $N_{\rm sample}$ & $p$  & $J_f$ & $\text{MSE}_{\rm CDF}^{\rm exact}$ & $\text{MSE}_{\rm CDF}^{\rm shots}$ & $\text{MSE}_{\rm PDF}^{\rm exact}$ & $\text{MSE}_{\rm PDF}^{\rm shots}$ & KL$_{\rm PDF}^{\rm shots}$ \\
  \hline
    Gamma        & $5 \cdot 10^4$   & $25$ & $2.9 \cdot 10^{-6}$ & $1\cdot10^{-5}$   & $1\cdot10^{-5}$ & $9\cdot10^{-4}$ & $2\cdot 10^{-3}$ & $4\cdot 10^{-3}$\\
    Gaussian mix & $2 \cdot 10^5$   & $30$ & $4.4 \cdot10^{-6}$  & $9\cdot 10^{-6}$  & $9\cdot10^{-6}$ & $2 \cdot 10^{-3}$ & $4\cdot 10^{-3}$ & $6\cdot 10^{-3}$\\
    $t$          & $5\cdot 10^4$    & $20$ & $2.1 \cdot 10^{-6}$ & $3\cdot 10^{-4}$  & $3\cdot10^{-4}$ & $1 \cdot 10^{-3}$ & $3 \cdot 10^{-3}$ & $5\cdot 10^{-3}$\\
    $s$          & $5\cdot 10^4$    & $20$ & $7.9 \cdot 10^{-6}$ & $3\cdot 10^{-4}$  & $3\cdot10^{-4}$ & $1 \cdot 10^{-2}$ & $1 \cdot 10^{-2}$ & $4\cdot 10^{-3}$\\
    $y$          & $5\cdot 10^4$    & $8$  & $3.7 \cdot 10^{-6}$ & $3\cdot 10^{-4}$  & $3\cdot10^{-4}$ & $9 \cdot 10^{-4}$ & $2 \cdot 10^{-3}$ & $2\cdot 10^{-3}$\\
  \hline \hline
  \end{tabular}
  \caption{Summary of the results. From left to right, we show the number of points 
  of the datasets, the number of variational parameters of the adiabatic evolution models,
  the best loss function value registered during the training, the MSE metric values
  respectively comparing the CDF estimation in exact simulation, the CDF estimation 
  in shot-noise simulation, the PDF estimation in exact simulation and the PDF 
  estimation in shot-noise simulation. When considering shot-noise simulation, 
  each estimation contributing to the MSE has been calculated as mean value of 20
  estimates, each of these obtained collecting $N_{\rm shots}=2\cdot 10^5$ shots 
  of the quantum circuit. The MSE valued for the Gamma and the Gaussian mixture
  examples are calculated using five hundred points equispaced in $[0,1]$, while 
  the HEP results are calculated considering the values extracted from a histogram 
  representation of the data setting $N_{\rm bins}=34$. In the last column, we 
  show the value of the Kullback-Leibler divergence between the estimated 
  distribution in case of $N_{\rm shots}=2\cdot 10^5$ and the theoretical distribution.}
  \label{tab:summary}
  \end{table*}

\begin{figure*}[ht]
  \centering
  \includegraphics[width=0.48\linewidth]{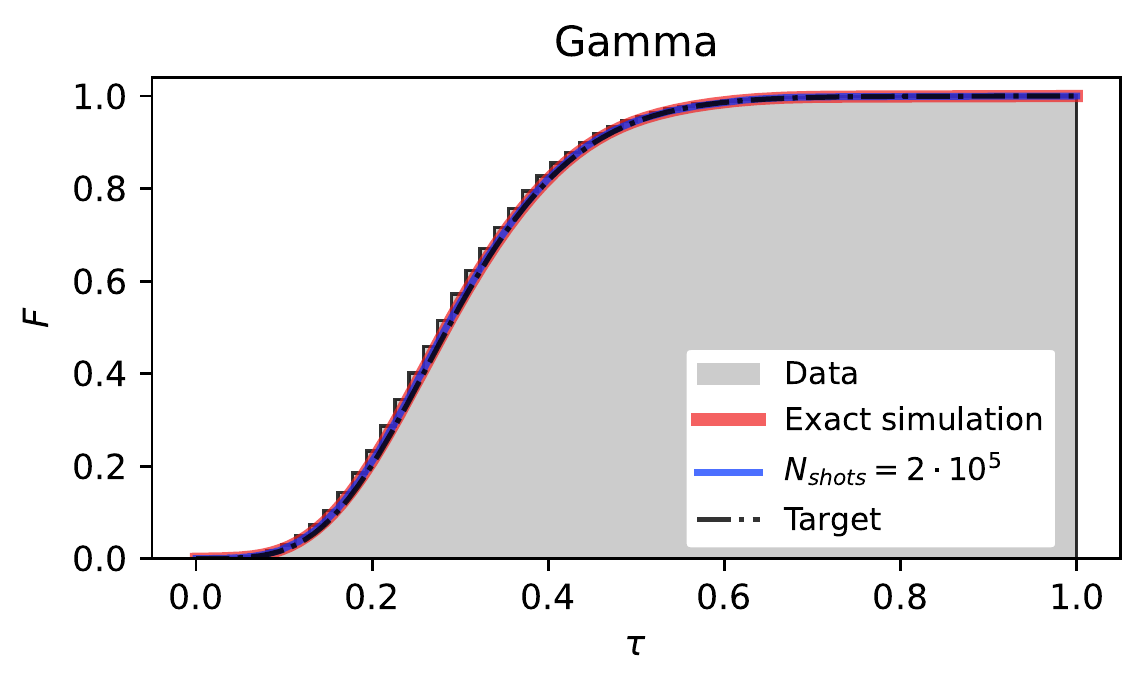}%
  \,\,
  \includegraphics[width=0.48\linewidth]{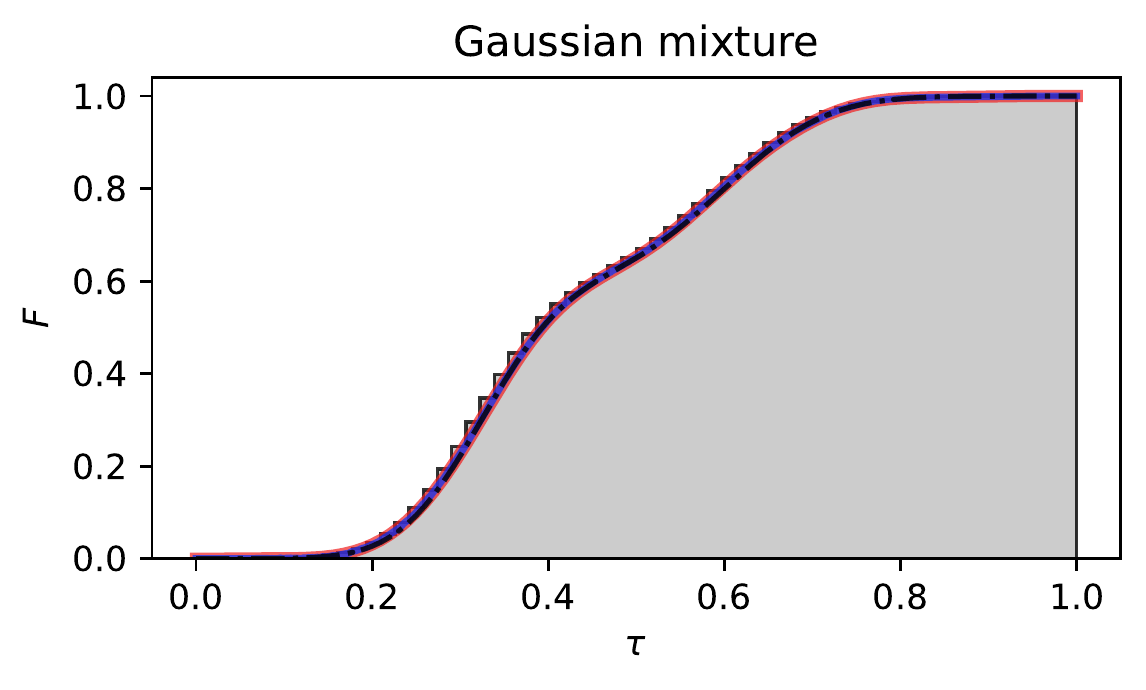}
  \includegraphics[width=0.5\linewidth]{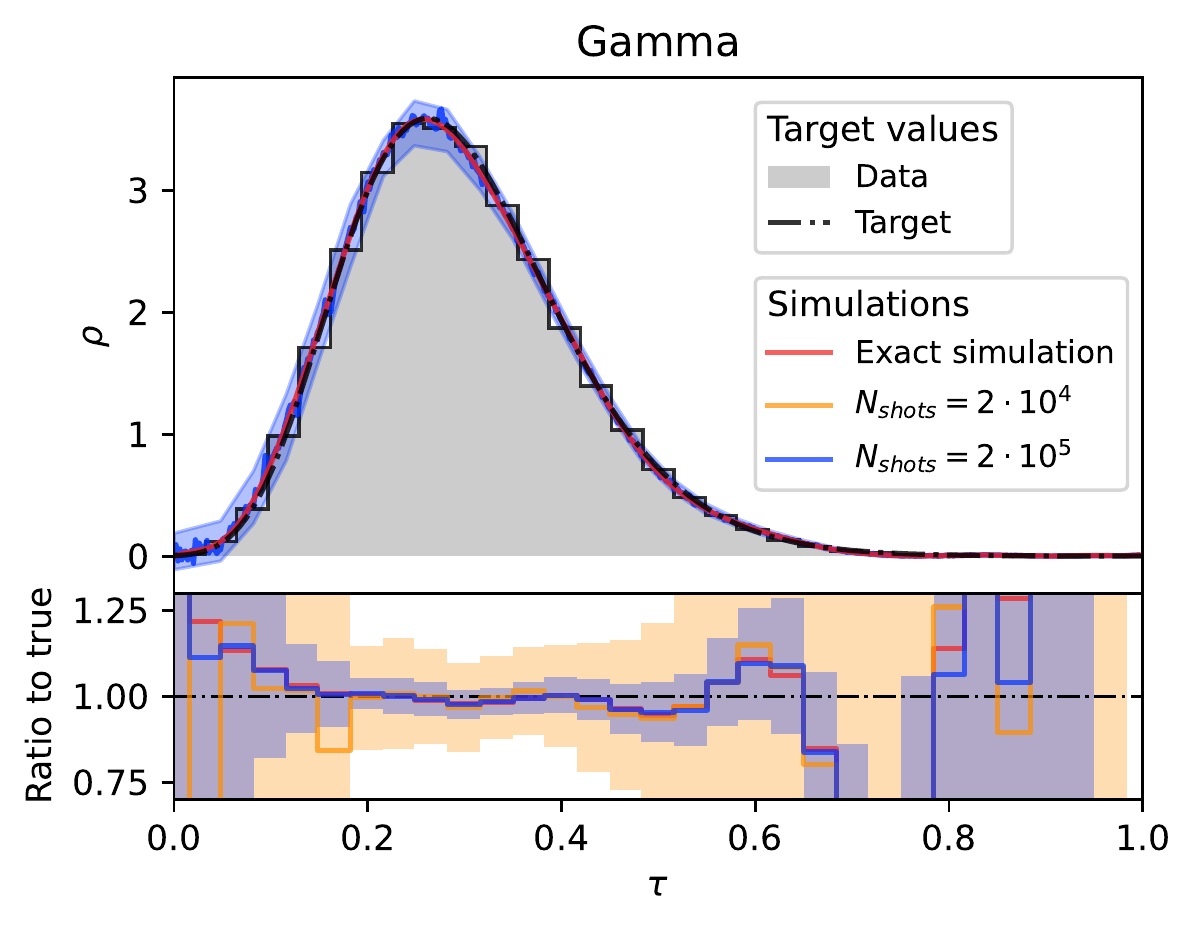}%
  \includegraphics[width=0.5\linewidth]{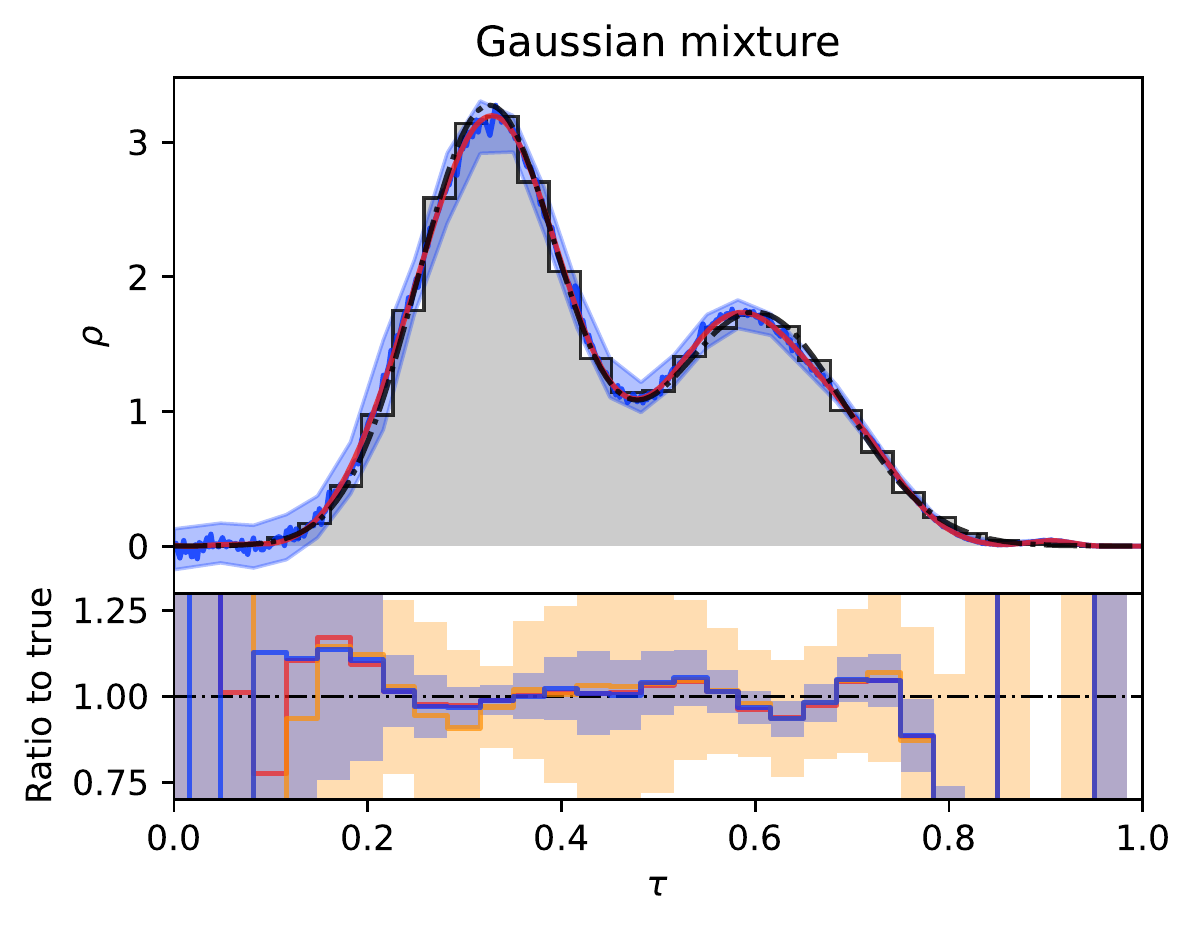}
  \caption{Top row: CDF fit via QAML procedure. The target CDF (dashed black line)
  is compared with QAML predictions obtained via exact simulation (red line)
  and simulation with shot-noise (blue line). An histogram of the data is shown in 
  grey. Bottom row: comparison between the target PDF and the
  result of the derivative of the trained circuit. Once again the target labels (dashed black line)
  are compared with the data histogram (grey),
  with QAML exact simulation results (red line) and QAML shot-noise simulations with $N_{\rm shots} 
  = 2\cdot 10^4$ and $N_{\rm shots} = 2\cdot 10^5$ (yellow and blue lines respectively).
  These same results are also shown in the form of a ratio between the target labels and the
  predictions via QAML in the lowest part of the figures. While representing the PDFs,
  only one simulation with shot-noise is drawn ($N_{\rm shots} = 2\cdot 10^5$).
  All the shot-noise simulations curves are represented together with
  a $1\sigma$ confidence belt calculated repeating $N_{\rm runs}=20$ the predictions 
  with fixed trained model.}
  \label{fig:cdf_examples}
\end{figure*}

\begin{figure*}[ht]
  \centering
    \includegraphics[width=0.3\linewidth]{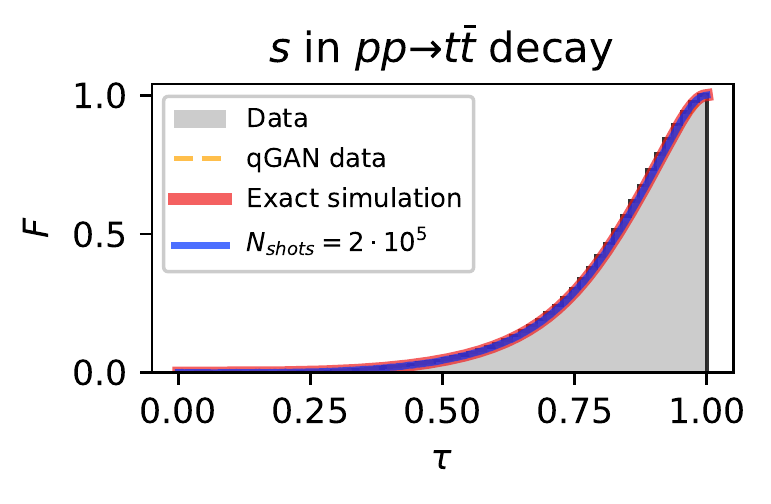}%
    \includegraphics[width=0.3\linewidth]{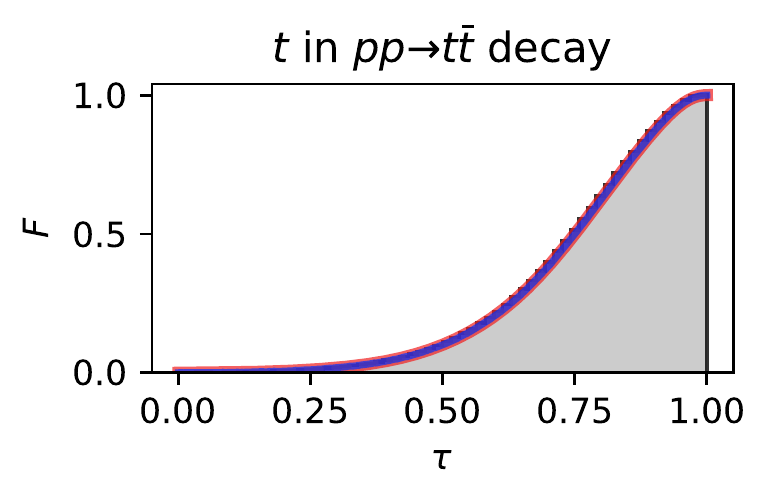}%
    \includegraphics[width=0.3\linewidth]{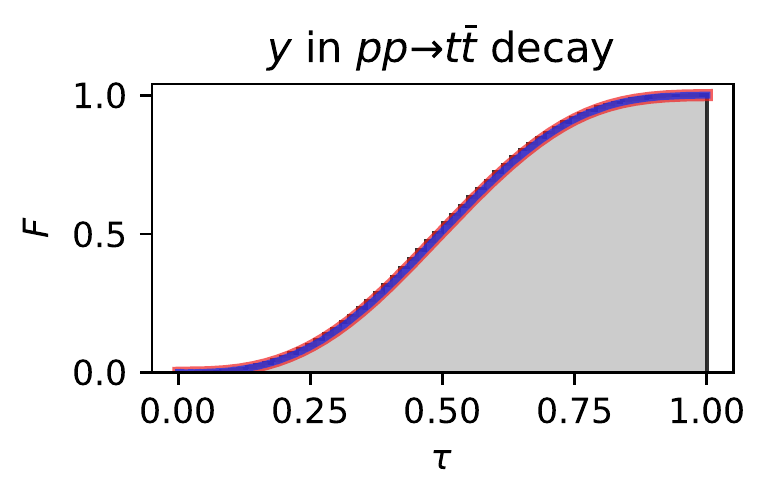}
    \includegraphics[width=0.3\linewidth]{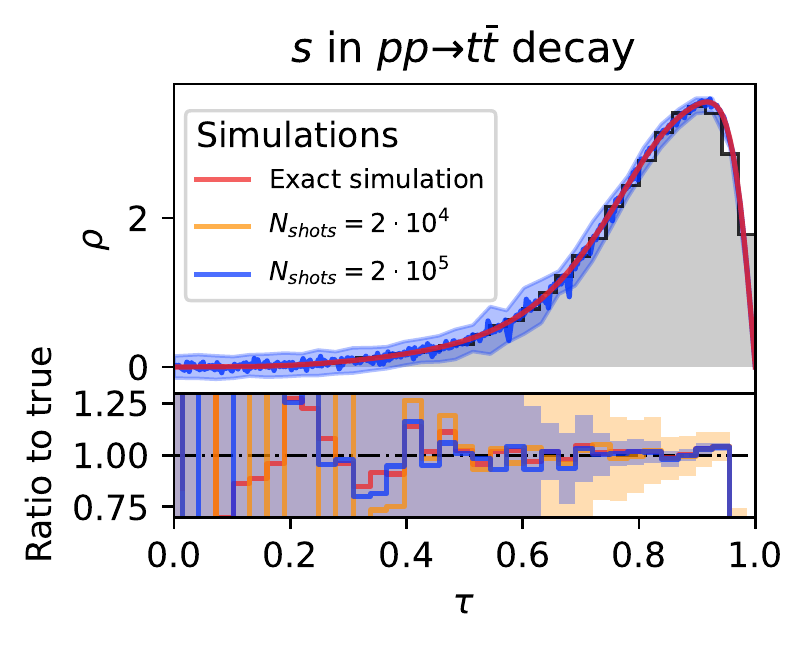}%
    \includegraphics[width=0.3\linewidth]{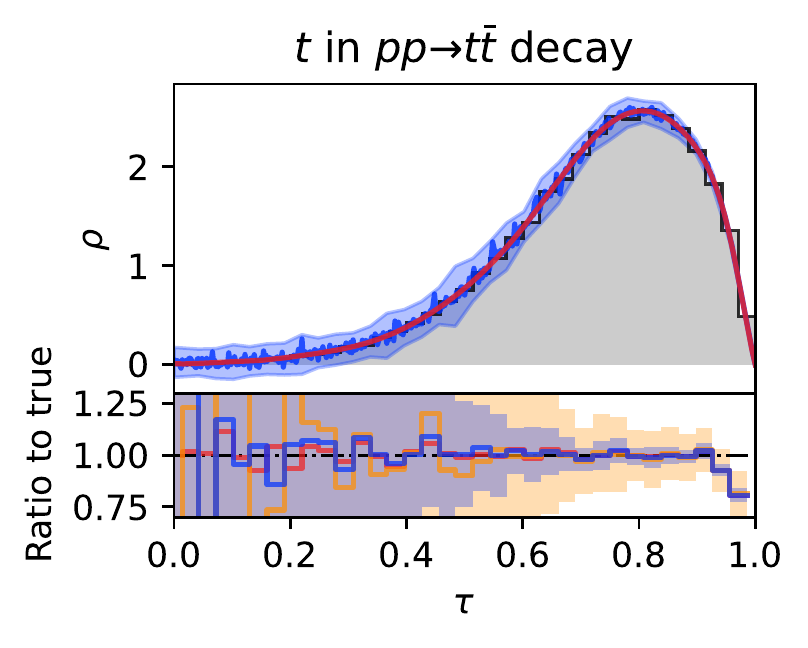}%
    \includegraphics[width=0.3\linewidth]{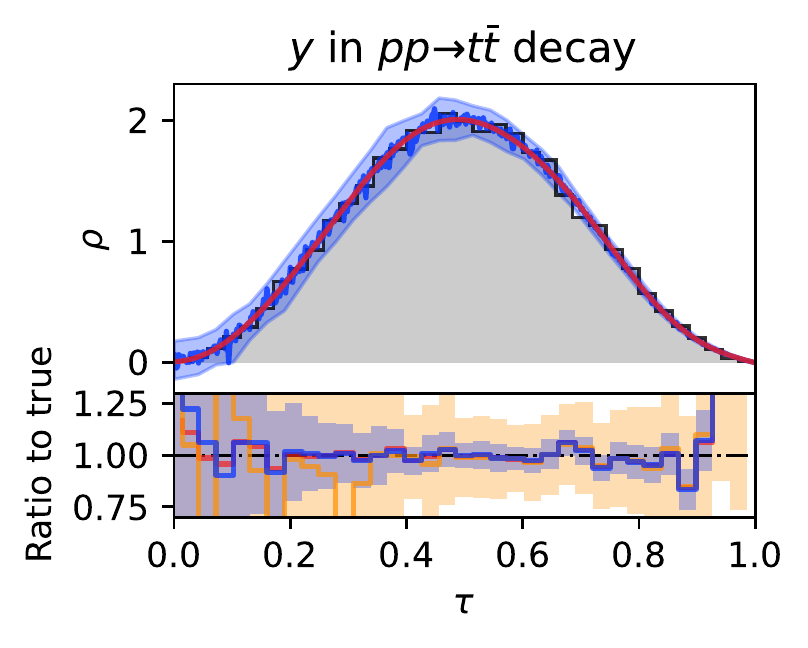}
    \caption{Top row: CDF fit via QAML procedure considering the HEP targets. 
    The target CDF (dashed black line)
    is compared with QAML predictions obtained via exact simulation (red line)
    and simulation with shot-noise (blue line). An histogram of the data is shown in 
    grey. Bottom row: comparison between the target PDF and the
    result of the derivative of the trained circuit. Once again the target labels (dashed black line)
    are compared with the data histogram (grey),
    with QAML exact simulation results (red line) and QAML shot-noise simulations with $N_{\rm shots} 
    = 2\cdot 10^4$ and $N_{\rm shots} = 2\cdot 10^5$ (yellow and blue lines respectively).
    These same results are also shown in the form of a ratio between the target labels and the
    predictions via QAML in the lowest part of the figures. While representing the PDFs,
    only one simulation with shot-noise is drawn ($N_{\rm shots} = 2\cdot 10^5$).
    All the shot-noise simulations curves are represented together with
    a $1\sigma$ confidence belt calculated repeating $N_{\rm runs}=20$ the predictions 
    with fixed trained model.}
    \label{fig:cdf_hep}
  \end{figure*}

In order to validate and test the procedure, we select two known distributions: 
a Gamma distribution and a Gaussian mixture of two Gaussian distributions. 
For each case, we generate a representative sample of dimension
$N_{\rm sample}$ and fit the resulting empirical CDF using the approach
described in Sec.~\ref{sec:regressor}. We then derive the PDF with the procedure 
introduced in Sec.~\ref{sec:cdf2pdf} and compare the results with the original 
distribution. We repeat this exercise for every
example by using quantum simulation on classical hardware with exact
state-vector representation and with shot-noise.

In these examples the adiabatic evolution training is set to stop once a given threshold value $J_{\rm thresh}$
of the loss function~\eqref{eq:loss_function} is reached. 
Furthermore, every sample is rescaled to be between $0$ and $1$ according 
to the definition of $\tau$ introduced in Sec.~\ref{sec:training}. In
all cases the adiabatic evolution is run from $\tau=0$ to $\tau=1$ with $\dd \tau=0.002$.
We define the scheduling function as a polynomial of order $p$
following the ansatz in Eq.~\eqref{eq:scheduling_ansatz}.

We start by drawing samples from a Gamma distribution, defined as
\begin{equation}
    \rho(x; \alpha, \beta) = \frac{\beta^{\alpha} x^{\alpha - 1} e^{\,-\beta x}}{\Gamma(\alpha)}, 
    \label{eq:gamma}
\end{equation}
with $\alpha = 10$ and $\beta = 0.5$. We take $N_{\rm samples}=5\cdot 10^{4}$
points and train the scheduling function until a target precision of
$J_{\rm thresh}= 10^{-5}$ is reached.

We repeat the procedure both with and without shot-noise.
In the case of simulations with shot-noise, we construct the final value of our 
predictions by repeating the calculation of the expectation value of $\sigma_z$ twenty times. 
The collected results are then used to define the prediction and its uncertainty 
as the mean and standard deviation of the obtained values, respectively. The discussed 
results are shown in Tab.~\ref{tab:summary}.

The results of the training can be seen in the first row and left column of
Fig.~\ref{fig:cdf_examples}, where we plot the empirical CDF (black line) together with both the
exact (red line) and shot-noise (blue and yellow lines) simulation, and a histogram of the data.
In the second row and left column of Fig.~\ref{fig:cdf_examples} we show the PDF
obtained by taking the derivative of the circuit compared it with the original
distribution, as well as the ratio with respect to the target labels.
We find that while the exact simulation achieves an accuracy of a few percent everywhere but the tails of the distribution, when shot-noise is enabled it is necessary to go beyond $N_{\rm shots}=10^{5}$ to achieve single-digit precision.
This is illustrated by the two shadowed bands for $N_{\rm shots}=2\cdot 10^5$ (blue), and 
a lower value of $N_{\rm shots}=2\cdot10^4$ (yellow) which has a worse than 10\% precision everywhere.

% JCM: I don't understand this sentence?
% To reconstruct the original dimensionality, the derivative of the target observable has to be rescaled multiplying it with the final real-time value $T$.

In order to test the algorithm with a more complicated example we also sample from 
a Gaussian mixture
\begin{equation}
  \rho(x; \vec{\mu}, \vec{\sigma}) =  0.6\mathcal{N}(x; \mu_1, \sigma_1)
  + 0.4\mathcal{N}(x; \mu_2, \sigma_2) \\,
  \label{eq:gaussian_mixture}
\end{equation}
with $\vec{\mu}=(-10, 5)$ and $\vec{\sigma}=(5, 5)$. From this
distribution we take $N_{\rm sample}=5\cdot 10^{5}$ points to generate the training
sample.
The results corresponding to this second target are shown in the right column of 
Fig.~\ref{fig:cdf_examples} and follow the same graphical conventions presented above. 

To quantify the accuracy of our predictions we define a Mean Squared Error (MSE)
metric
\begin{equation}
\text{MSE} = \frac{1}{N}\sum_{j=0}^{N} (y_{j, \rm meas} - y_{j, \rm pred})^2 \\,
\label{eq:mse}
\end{equation}
where $y_{j, \rm meas}$ and $y_{j, \rm pred}$ correspond to the target label and 
the model prediction associated to a specific variable $x_j$. The Eq.~\eqref{eq:mse}
is written as function of a general variable $y$, which is then deployed as the CDF and 
the PDF predictions in what follows. In addition, we also compute the Kullback-Leibler
divergence
\begin{equation}
\text{KL} = \sum_{i=1}^N  \biggl[ y_{i, \rm pred} \, \log \biggl(\frac{y_{i, \rm pred}}{y_{j, \rm meas}}\biggr) \biggr]
\label{eq:kl}
\end{equation} 
for the PDFs computed in the case of shot-noise simulation with $N_{\rm shots}=2\cdot 10^5$ shots.
We collect our results in Tab.~\ref{tab:summary}.

\subsection{Density estimation of simulated high energy physics data}

In the previous case we were training the circuit using a sample from a
known distribution, we now address the more complex case of learning an unknown
distribution.
We consider the particle physics process involving top and
anti-top quark pair production ($pp \to t\bar{t}$), 
for which we choose the cross section differential on the rapidity $y$
and the logarithms of the Mandelstam variables $-\log{(-t)}$ and $-\log{s}$.
This choice is motivated by the following reasons: on one hand, they present a real-world scenario in high energy physics (HEP), stress-testing the method, and on the other hand, it provides a potential use-case for the methods presented in this paper.

While in general one would train directly on the output of a Monte Carlo event generator, in our test case the data sampling is obtained from a separated hybrid classical-quantum model called Style-qGAN~\cite{Bravo_Prieto_2022}.
The Style-qGAN has been trained with $10^5$ events for $pp\to t\bar{t}$
production at a center of mass of $\sqrt{s}=13$ TeV computed at Leading Order.

In Fig.~\ref{fig:cdf_hep} we show the results for the training of the circuit
(the CDF on the top row) and its derivative (the PDF on the bottom row),
following the same graphical conventions introduced in the previous section.

In Table~\ref{tab:summary} we summarize the results for all examples
tested in this section. For each model we describe the final configuration and
fit accuracy by calculating the MSE values for both the CDF and the PDF estimations.

We find the achieved level of quality is satisfactory for all tested
distributions. We also observe that $N_{\rm shots}=2\cdot 10^{5}$ shots provide
sufficient statistics to achieve a precision in the range of 4-10\% (see ratio plots in the bottom part of Fig.~\ref{fig:cdf_examples} and Fig.~\ref{fig:cdf_hep}).
In the case of the HEP based data, this level of precision is only achieved near the distribution peaks.
The precision (and the accuracy) deteriorates as we move towards the tails where the differential cross section approaches zero, a behavior consistent with classical approaches a common challenge when fitting distributions~\cite{Butter:2019cae}.

\begin{figure*}[ht]  
  \includegraphics[width=0.31\linewidth]{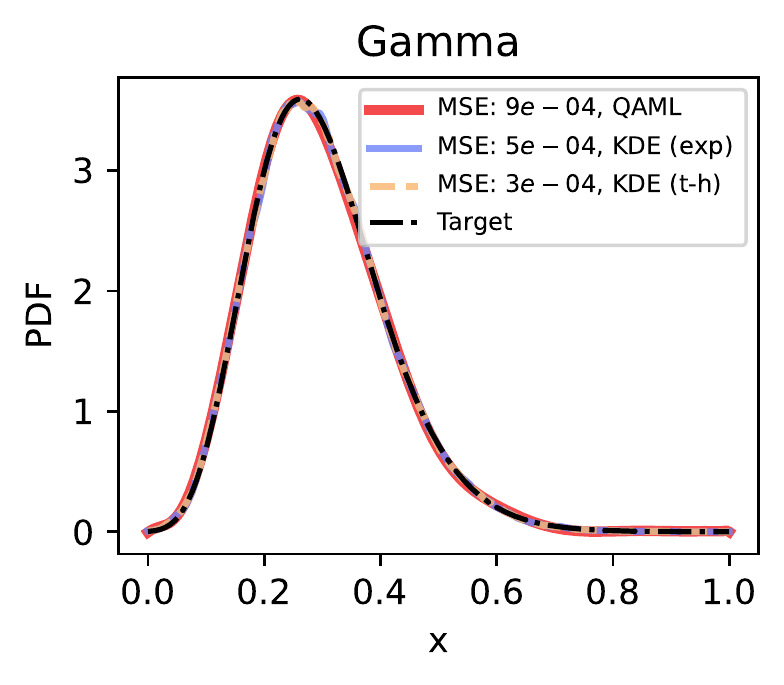}%
  \includegraphics[width=0.31\linewidth]{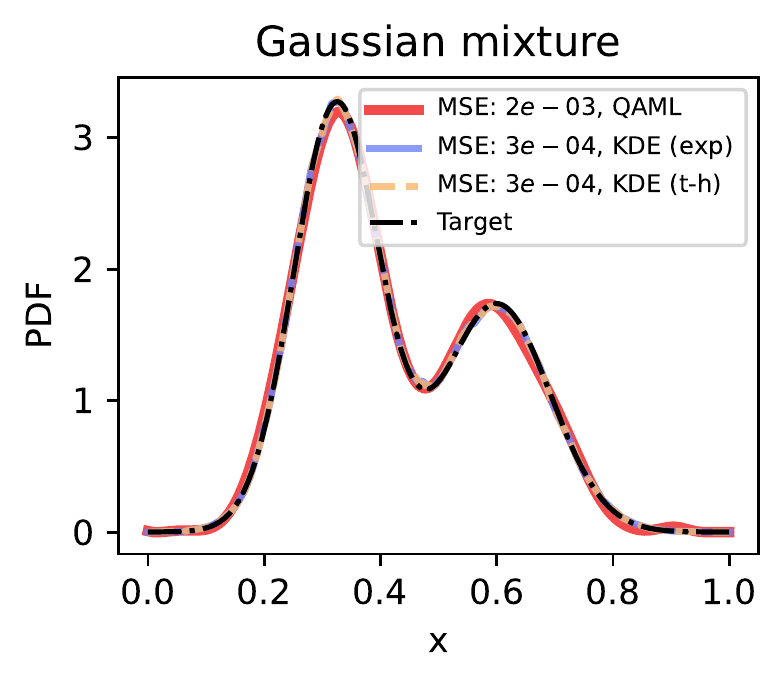}%
  \includegraphics[width=0.325\linewidth]{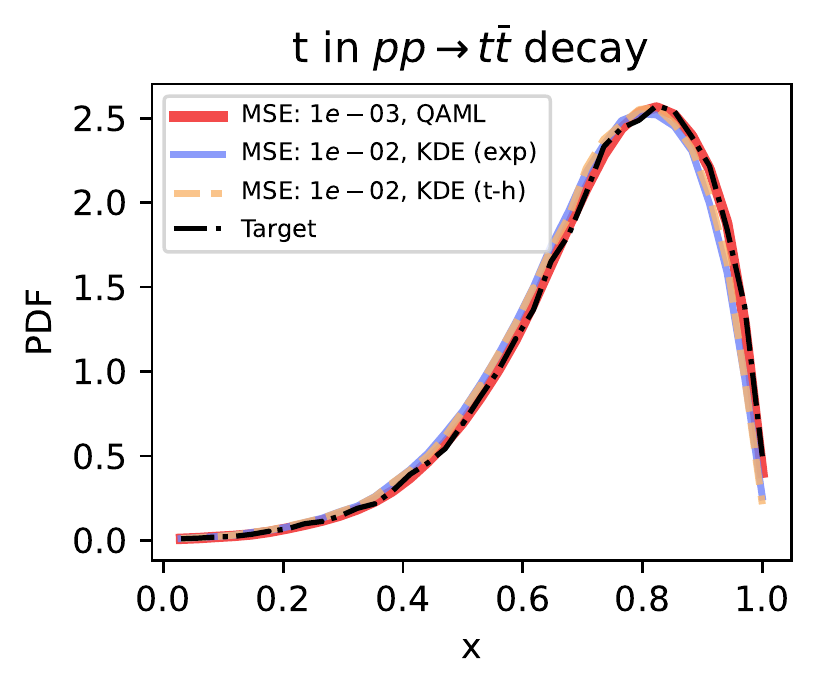}
  \caption{\label{fig:kde} Comparison between our QAML trained results (red lines)
  and the proposed solutions of 1D Kernel Density Estimation method provided 
  by \texttt{scikit-learn} using \textit{top-hat} (orange lines) and \textit{exponential} 
  (blue lines) kernels. All the approximations are compared with the 
  target values (black lines).}
\end{figure*}

\subsection{Benchmark with KDE methods}

In this section we benchmark our results with a state-of-art Kernel Density Estimation 
method provided by \texttt{scikit-learn}~\cite{scikit-learn}.
While we are optimistic about 
the future of quantum technologies and the potential utility of the proposed method, we stress that 
this analysis is intended here to only verify the robustness of the results in terms of accuracy,
and not to propose this method as an alternative to classical techniques. 
This kind of stress-testing is necessary in order to have confirmation of the 
validity of the proposed method once faster and more accurate hardware is obtained. 

We compare our exact-simulation results with the predictions computed using the \texttt{sklearn.neighbours.KernelDensity}
method selecting \textit{top-hat} and \textit{exponential} kernels. The first has 
been chosen for its computational efficiency, while the second is more suitable 
when dealing with tailed distributions. 
The kernel bandwidth has been hyper-optimized using
a Tree of Parzen Estimators (TPE) on a random grid of one thousand points sampled 
from the interval $[10^{-5}, 1]$. The hyper-optimization has been performed using 
\texttt{hyperopt}~\cite{bergstra2015hyperopt}. 

We compute this benchmark in order to check whether our algorithm 
is competitive in terms of accuracy with respect to the most commonly used classical 
tools. We test both the QAML procedure and the KDE algorithm on the Gamma distribution, 
the Gaussian mixture and the $t$ Mandelstam variable, which is the one corresponding
to our best result in the HEP examples. We show these results in Fig.~\ref{fig:kde},
where our predictions are promisingly close to those obtained through KDE estimation. 
Additionally, in the case of the variable $t$, 
which is defined on a more critical domain due to the chosen binning, QAML seems to offer a 
more accurate fit to the density profile.

Considering the execution time of the two algorithms, they present very different 
characteristics: on one hand, QAML suffers the 
optimization time, which is necessary to perform the density estimation. On the other 
hand, when the training is concluded, computing the PDF is extremely lightweight.
Instead, in the case of the KDE, the opposite considerations can be made:
the hyper-optimization is relatively light, but then the PDF evaluation depends on the bandwidth size.

\section{Hardware}
\label{sec:hardware}

In order to assess the performance of the model in real quantum hardware
we use a 5-qubits superconducting chip hosted in the Quantum Research Centre (QRC) of the 
Technology Innovation Institute (TII).
We calculate the predictions for the values of the CDF
using the best parameters obtained through the training with the shot-noise 
simulation whose results are presented in Tab.~\ref{tab:summary}. We take into
account the Gamma distribution defined in Eq.~\eqref{eq:gamma} and we do not 
apply error mitigation techniques to the hardware, in order to explore the 
potentialities of the bare chips. 

We consider $N_{\rm data}=25$ points equally distributed in the target range $[0,1]$
and for each of these we perform $N_{\rm runs}=10$ predictions executing $N_{\rm shots}=1000$
times the circuit on the quantum hardware. Using these data, we calculate the 
final predictors and their uncertaintes as mean and standard deviation over the 
$N_{\rm runs}$ results. We also calculate the MSE introduced in Eq.~\eqref{eq:mse}
to evaluate the fit accuracy.

\begin{figure}[ht]  
  \includegraphics[width=1\linewidth]{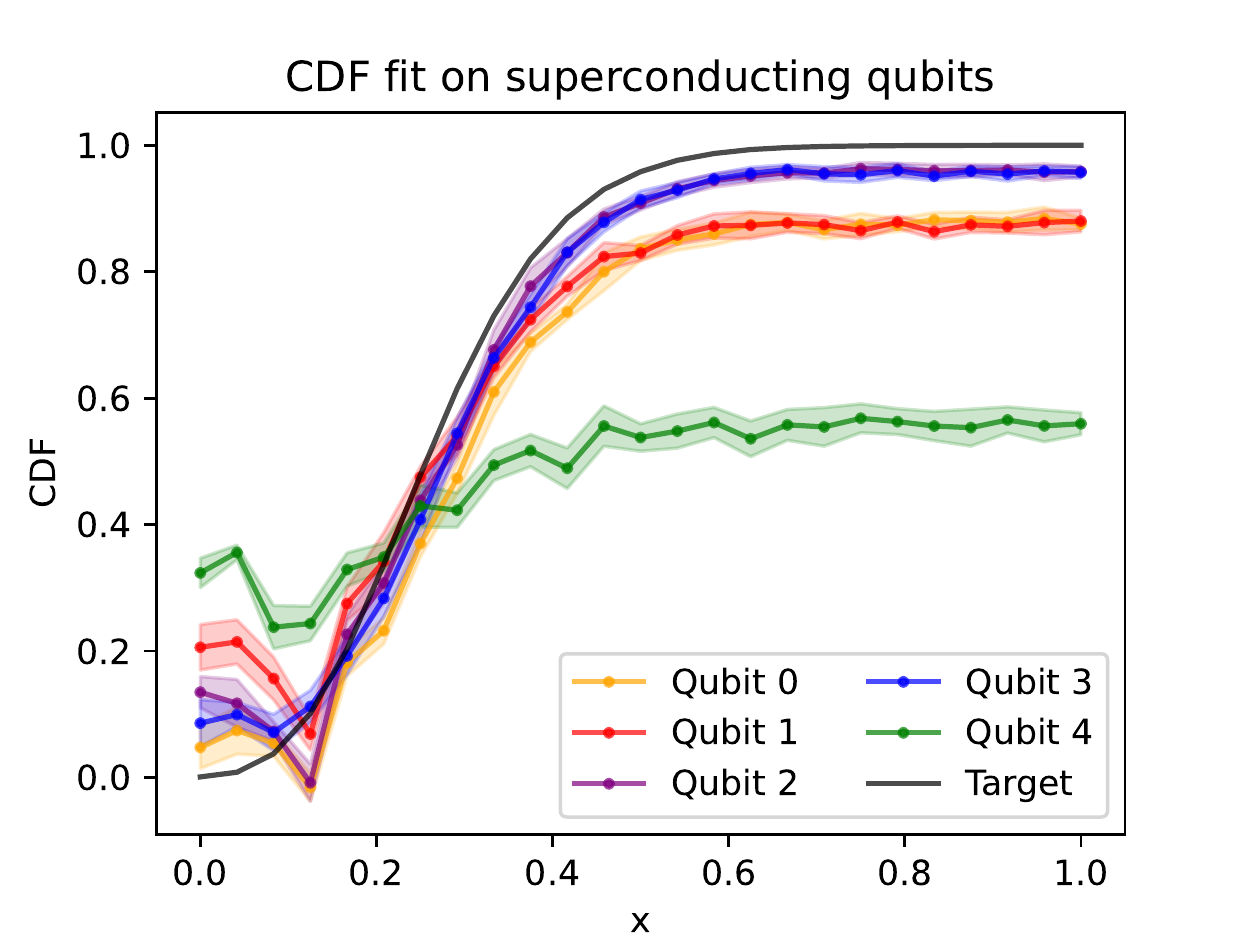}
  \caption{$N_{\rm runs}=10$ predictions are performed for $N_{\rm data}=25$ points
  in the target range $[0,1]$ using each qubit of a 5-qubits device hosted in the 
  QRC.}
  \label{fig:predictions_on_hardware}
\end{figure}

In order to study how the CDF predictions deteriorate when executed in hardware
and to study the dependence of this deterioration on how the qubits are tuned,
we repeat this procedure for each of the five qubit of the device. 
The results are shown in Fig.~\ref{fig:predictions_on_hardware} and are in 
agreement with Tab.~\ref{tab:qubits_summary}, where we report the assignment 
fidelities~\cite{gao2021practical} of the qubits and the calculated MSE values.

\begin{table}[ht]
\begin{tabular}{ccc}
\hline \hline
Qubit ID &  Assignment Fidelity  & MSE \\
\hline 
$0$ & $0.926$ & $1.3\cdot10^{-2}$ \\
$1$ & $0.886$ & $1.4\cdot10^{-2}$ \\
$2$ & $0.953$ & $3.4\cdot10^{-3}$ \\
$3$ & $0.952$ & $2.7\cdot10^{-3}$ \\
$4$ & $0.707$ & $1.3\cdot10^{-1}$ \\
\hline \hline
\end{tabular}
\caption{Prediction deployed on superconducting chip. For each qubit of the device 
we show the assignment fidelity at the moment of the execution and the MSE values.}
\label{tab:qubits_summary}
\end{table}

The quantum hardware control is performed using \texttt{Qibolab}~\cite{Efthymiou2024qibolabopensource, 
stavros_efthymiou_2023_7748527} 
and the qubits are characterized and calibrated executing the \texttt{Qibocal}'s 
routines~\cite{pasquale2023opensource, andrea_pasquale_2023_7662185}.  

\section{Conclusion}
\label{sec:conclusion}

In this work we presented a proof-of-concept application of quantum machine learning which makes 
use of both analog and gate-based quantum computation addressing different 
tasks within the same problem. Namely, we introduced an algorithm for the determination of probability
density functions. We first define a mechanism
to use adiabatic evolution as a regression model for the fit of the
cumulative density function of a sample. The adiabatic evolution model is then 
represented by the Trotterization of the adiabatic Hamiltonian in terms of a quantum circuit. 
The probability density profile is then calculated by applying parameter shift rules to the obtained circuit. 
This method allows the usage for training and inference of quantum devices designed 
for both annealing and circuit-based technologies. The numerical results obtained and presented in
Sec.~\ref{sec:validation} show successful applications of the methodology for
predefined PDFs and empirical distributions obtained from high-energy particle
physics observables. In the same section we compare the presented methodology
with classical Kernel Density Estimations techniques, demonstrating that the proposed 
method can yield satisfactory results when compared to state-of-art algorithms. 
Finally, in Sec.~\ref{sec:hardware} we deployed the trained models on to a superconducting
bare device, showing interesting results even without applying any quantum 
error mitigation algorithm. 

All numerical results have been obtained using {\tt Qibo}~\footnote{\url{https://github.com/qiboteam/qibo}}, 
a full-stack and open source framework for quantum computing, and are publicly
available in~\footnote{\url{https://github.com/qiboteam/adiabatic-fit}}.
Further possible developments include the generalization of this method for the
simultaneous determination of multi-dimensional probability density distributions, the
deployment of the full training procedure on quantum devices and the possibility
to use real quantum annealing for the optimization of the regressing model
parameters. \\

\acknowledgments SC thanks the TH hospitality during the elaboration of this manuscript. \\

\noindent
\textbf{Funding:} 
This project is supported by CERN's Quantum Technology Initiative (QTI). 
MR is supported by CERN doctoral program. \\

\noindent
\textbf{Author contributions:} 
All authors contributed equally to the elaboration of this manuscript.\\

\noindent
\textbf{Data Availability:} 
The code to reproduce the simulations can be found at~\cite{adiabatic-fit}. 

\section*{Declarations}

\noindent
\textbf{Conflict of interest:}
The authors declare no conflict of interest.\\

\noindent
\textbf{Open access:}
This article is licensed under a Creative Commons Attribution 4.0 International 
License, which permits use, sharing, adaptation, distribution and reproduction 
in any medium or format, as long as you give appropriate credit to the original 
author(s) and the source, provide a link to the Creative Commons licence, 
and indicate if changes were made. The images or other third party material in 
this article are included in the article's Creative Commons licence, unless 
indicated otherwise in a credit line to the material. 
If material is not included in the article's Creative Commons licence and your 
intended use is not permitted by statutory regulation or exceeds the permitted 
use, you will need to obtain permission directly from the copyright holder. 
To view a copy of this licence, visit
\url{https://creativecommons.org/licenses/by/4.0/}.

\bibliography{references}

\end{document}